\documentclass[lettersize,conference]{IEEEtran}

\ifCLASSINFOpdf
  \usepackage[pdftex]{graphicx}
  \graphicspath{{./}{../pdf/}{../jpeg/}}
  \DeclareGraphicsExtensions{.pdf,.jpeg,.png}
  \usepackage[caption=false, font=footnotesize]{subfig}
\else
\fi

\usepackage{amsmath}
\usepackage[cmintegrals]{newtxmath}

\usepackage{algorithmicx}
\usepackage{algorithm}
\usepackage{algpseudocode}
\makeatletter
\newcommand{\removelatexerror}{\let\@latex@error\@gobble}
\makeatother

\usepackage{cite}

\usepackage{tabularx}
\usepackage{booktabs}

\usepackage{color,soul}
\setulcolor{red}
\sethlcolor{yellow}
\usepackage{xcolor}

\renewcommand\hl[1]{#1} 
\usepackage{makecell}

\begin{document}

\title{Toward Multi-Layer Networking for Satellite Network Operations}


\author{\IEEEauthorblockN{
Peng Hu\IEEEauthorrefmark{1}\IEEEauthorrefmark{2}}
\IEEEauthorblockA{\IEEEauthorrefmark{1}
Dept. of Electrical and Computer Engineering, University of Manitoba, Winnipeg, Canada}
\IEEEauthorblockA{\IEEEauthorrefmark{2}David R. Cheriton School of Computer Science, University of Waterloo, Waterloo, Canada}
Email: peng.hu@umanitoba.ca
}

\markboth{Journal of \LaTeX\ Class Files,~Vol.~00, No.~0, 00~2024}%
{Shell \MakeLowercase{\textit{et al.}}: Bare Demo of IEEEtran.cls for IEEE Journals}

\maketitle
 
\begin{abstract}
Recent advancements in low-Earth-orbit (LEO) satellites aim to bring resilience, ubiquitous, and high-quality service to future Internet infrastructure. However, the soaring number of space assets, increasing dynamics of LEO satellites and expanding dimensions of network threats call for an enhanced approach to efficient satellite operations. To address these pressing challenges, we propose an approach for satellite network operations based on multi-layer satellite networking (MLSN), called ``SatNetOps''. Two SatNetOps schemes are proposed, referred to as LEO-LEO MLSN (LLM) and GEO-LEO MLSN (GLM). The performance of the proposed schemes is evaluated in 24-hr satellite scenarios with typical payload setups in simulations, where the key metrics such as latency and reliability are discussed with the consideration of the Consultative Committee for Space Data Systems (CCSDS) standard-compliant telemetry and telecommand missions. Although the SatNetOps approach is promising, we analyze the factors affecting the performance of the LLM and GLM schemes. The discussions on the results and conclusive remarks are made in the end. 
\end{abstract}


%
\IEEEpeerreviewmaketitle

\section{Introduction}
%
%
%
%
With the recent advancements and deployments of low-Earth-orbit (LEO) satellites, the upcoming space assets will approach a unprecedented volume in the coming years. These space assets, while providing much convenience and resiliency for global telecommunications networks, demand efficient, reliable, and robust operations for the satellite networks in ``New Space'' \cite{Paikowsky2017} ecosystems.

\begin{figure}[!t]
\centering
\includegraphics[width=\linewidth]{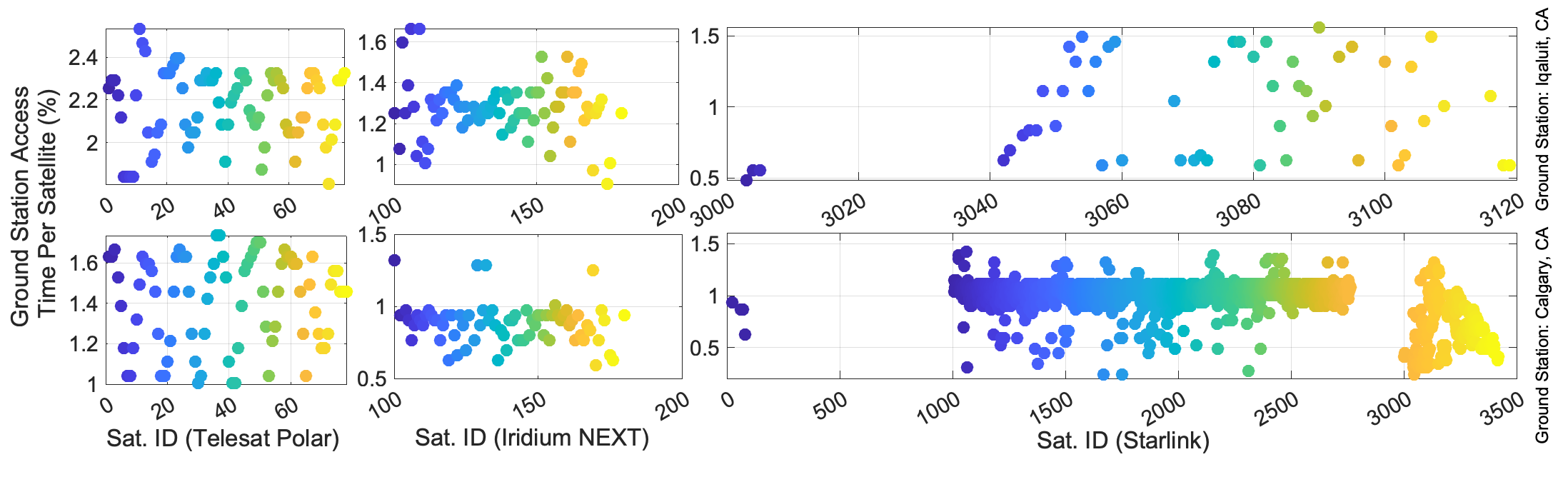}
\caption{Access analysis of typical LEO satellite constellations}
\label{Fig:sat_access}
\end{figure}

One of the essential services to support the new satellite operations is telecommand (TC) and telemetry (TM) missions, which rely on the efficient transmissions of packets for spacecraft operations. From a communications perspective, TC and TM missions share a similar process where TC messages are transferred from a ground station (GS) to a target satellite, while TM messages are transferred in the reverse direction and the packet transmission follow a similar pattern based on the Consultative Committee for Space Data Systems (CCSDS) standards. We therefore focus on the discussion of representative TC missions. In the traditional approach, TC messages are sent or received when a direct contact opportunity occurs between an operations GS and a target satellite. CCSDS \cite{CCSDS232.0-B-4} has standardized the transmissions of TC packets for space-to-space and ground-to-space communication links. The TC packets are usually small in size but require high reliability. However, such a traditional approach faces challenges that significantly reduce the efficiency of satellite operations. For example, the TC message transmissions depend on the access states. Based on our 24-hr access analysis of typical LEO satellite constellations shown in Fig. \ref{Fig:sat_access}, the average access opportunity per satellite in three typical constellations is very low for mostly $< 2\%$ chance of access to two typical GS locations (one in northern Canada and one in western Canada where most LEO satellites have coverage) with a minimum elevation 25$^{\circ}$ over a 24-hr mission, which clearly shows the limitations of the traditional operations approach in terms of efficiency and real-time communication capabilities. Although satellites are traditionally operated in isolation from their orbiting counterparts, communicating only with ground-based infrastructure with high delays leaves them susceptible to the consequences of various anomalies. Further, satellite networks have not been fully utilized for operations. These challenges call for efficient and resilient operations of a spacecraft and the entire satellite network.

Recently, using a legacy geostationary (GEO) satellite network for satellite operations missions has been considered. The constant positions of GEO satellites relative to ground components at a high altitude provide broad coverage of LEO satellites at a low altitude. For example, in November 2020, Inmarsat reportedly started a new service called Inter-satellite Data Relay System (IDRS), providing real-time links between GEO and LEO satellites. With the recent launch of I-6F1 GEO satellites with L/Ka-band payloads, new GEO satellites are expected to be deeply integrated into other medium-Earth-orbit (MEO) or LEO satellites into the 5G network as part of an ``ORCHESTRA'' network. Although the space industry shows intensive interest in an option of employing legacy or new satellite fleet for operations, there are no technical details or formal studies available about established schemes or assessments. In fact, such an option would face practical efficiency and timing challenges considering the distance between GEO and LEO. To reduce the distance for packet transmission, LEO satellite networks may be considered in combination with GEO satellites. The idea falls into the multi-layer satellite networking (MLSN) concept where satellites in different orbits are viewed in multiple layers. The layers may include various shells of a large constellation, different constellations, and different Earth orbits. However, MLSN has been mainly considered for data traffic routing, which differs from the TM/TC transmissions in practice for security and reliability considerations. The solutions to utilizing MLSN for operations missions are lacking.

This paper aims to address the new challenges of the operational needs faced by the recent growth of space assets in a timely manner. We propose a new approach considering MLSN, collectively called ``SatNetOps'', which can support efficient operations of non-geostationary (NGSO) satellite operations and enable the design of MLSN schemes. To the best of our knowledge, this is the first work addressing the new challenges for NGSO satellite network operations and providing a formal evaluation of MLSN-based schemes. The main contributions of the paper are summarized as follows:
\begin{itemize}
    \item We identify the pressing challenges of the current satellite operations and propose the ``SatNetOps'' approach with two generic schemes devised for TC missions, called GEO-LEO MLSN (GLM) and LEO-LEO MLSN (LLM).
    \item We evaluate the performance in terms of latency and reliability in different scenarios considering the feasible configurations. 
    \item We adopt the popular communications payloads in the recent satellites, i.e., the radio-frequency (RF) and free-space optical (FSO) communications technologies in space-to-space and space-to-ground communications.
\end{itemize}

The remainder of the paper is structured as follows. The related work is discussed in Section II. The proposed SatNetOps schemes for TC message transfer are discussed in Section III. The evaluation of the proposed schemes is discussed in Section IV. The conclusive remarks are presented in Section V.

\section{Related Work}
Compared to GEO satellites, throughput and latency are usually considered the two significant advantages of the current LEO satellite systems. The communications between layers of satellite networks can be linked to the concept of MLSN \cite{Nishiyama13}. Recent work and deployment plans of LEO satellite constellations are centered on throughput improvements. The multi-layer networking may occur through multiple shells of a large constellation with LEO satellites, such as Starlink's mega-constellation \cite{Pachler21}, although the cross-shell networking is considered complex \cite{Cakaj21}. Pachler \textit{et al.} \cite{Pachler21} showed the use of optical inter-satellite links (ISLs) on Telesat, Amazon Kuiper, and Starlink constellations can almost double the system throughput of each system compared to the non-optical ISLs. A cooperative communication multi-access scheme in MLSN was recently proposed in \cite{Ge2021}. For the LEO satellite operations, a real-world solution has recently realized by Inmarsat in late 2021 for using the GEO satellites for operations missions of LEO satellites and the SES also implied the benefits of using their upcoming o3b mPOWER fleet for satellite networking. However, MLSN schemes and the discussion of the related factors such as satellite and GS communications, MLSN strategies, and mission parameters are not available. 

CCSDS has developed the space data link protocols for TC missions which may be used with other upper layer CCSDS protocols. The latest issue of TC space data link protocol (TC-SDLP) released in October 2021 \cite{CCSDS232.0-B-4} supports two types of services: sequence-controlled (Type-A) and expedited (Type-B) services for missions with different priorities. Although both services have the same frame format, addressing, segmentation and blocking mechanisms, Type-A services support the Automatic Repeat Request (ARQ) mechanism for flow control, while Type-B services do not have flow control. Type-B services are used in ``exceptional operational circumstances'' such as spacecraft recovery, or flow control is provided at the upper layers \cite{CCSDS232.0-B-4}. Furthermore, these services in TC-SDLP only support unidirectional and asynchronous services where no predefined timing rules are specified. There is also no existing study on the recommendations for the timing performance for the operations of LEO satellite constellations. 
The timing performance of satellite communications needs to consider the recent developments of the satellite platforms. The performance of the TC missions also depends on the communication payload, where the L/S bands and Ka-band are broadly used. Since the first test of optical links for space missions in November 2014 \cite{Zech15}, FSO communication is expected to be well adopted by the space industry in the future. The recent tests initiated by the European Space Agency (ESA), the National Aeronautics and Space Administration (NASA) and the commercial LEO satellite constellations further indicate the increasingly planned use of FSO on space-to-space and ground-to-space links. However, the study considering these payloads for TC missions is lacking in the literature. 

\section{SatNetOps and System Model}

\subsection{SatNetOps}
\begin{figure}[!t]
\centering
\includegraphics[width=0.7\linewidth]{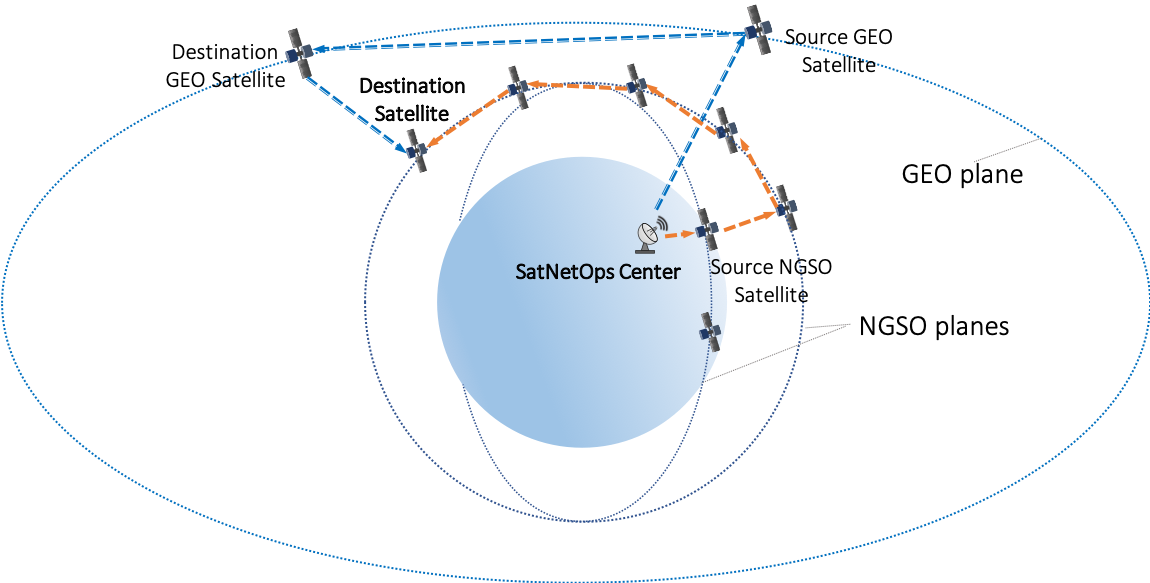}
\caption{Illustration of the proposed approach for SatNetOps.}
\label{Fig:satops_mln}
\end{figure}

The SatNetOps approach aims to enhance the efficiency of a number of spacecraft operations through satellite networking. SatNetOps intends to improve real-time and reliable communications for satellite operations. An example SatNetOps case is shown in Fig. \ref{Fig:satops_mln}, where a SatNetOps Center on Earth attempts to initiate a TC mission by sending a TC packet to a destination NGSO satellite. The packet traverses in multiple hops through one or more satellite networks, where the path established can also be used for transferring TM packets or control messages.

\subsection{Proposed SatNetOps Schemes}
\begin{figure}[!t]
\centering
\includegraphics[width=0.80\linewidth]{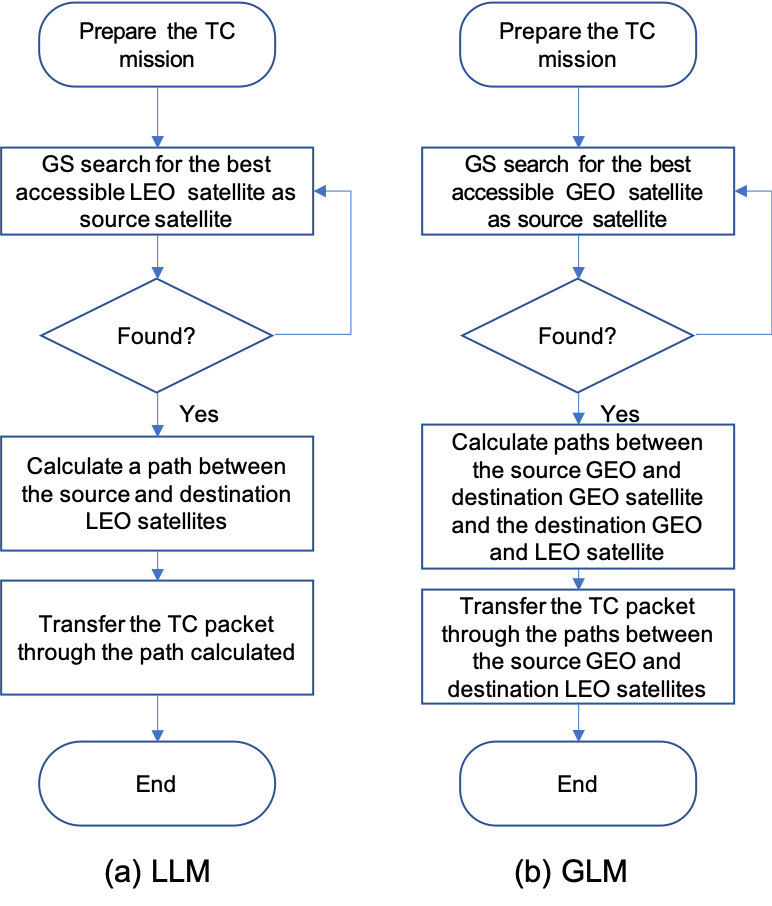}
\caption{Flowcharts for the proposed LLM and GLM schemes}
\label{Fig:flowchart} \vspace{-10pt}
\end{figure}

Here we propose two general schemes considering GEO and LEO satellites, called GLM and LLM. GLM uses GEO satellites to relay packets to the destination satellite, while LLM uses LEO satellites to forward packets to the destination satellite. GLM uses the GS-GEO, LEO-GEO, and GEO-GEO links, while LLM uses GS-LEO and LEO-LEO links. The brief descriptions of the schemes are shown in Fig. \ref{Fig:flowchart}. For both schemes, the TC mission preparation will determine the TC message, destination/target LEO satellite in a constellation, and the GEO satellites to be used. For LLM, the path calculation can occur dynamically due to constant satellite movements, where the inter- and intra-plane satellites are calculated based on the well-adopted configuration that each LEO satellite can access up to four neighbouring satellites in the same direction. Each LEO/GEO satellite will calculate a minimum elevation angle to ensure the next satellite is in line of sight when choosing the next hop. 


A SatNetOps scheme removes the need for direct contact times between a GS and a target satellite and ensures timing performance. Therefore, let us discuss the latency measure for GLM/LLM-based TC missions. The latency $D$ of a TC message transfer mission consists of four standard components: propagation delay, $D_{pa}$, transmission delay, $D_{t}$, processing delay, $D_{pc}$, and queuing delay, $D_{q}$. Since the TC packets are small, the average queuing delay $\bar{D}_{q}$ per satellite can be assumed as a constant, $m$, and the average $\bar{D}_{pc}$ per satellite assumed to be a constant, $k$. The size of the TC packet in bytes is $M$. The speed of light in FSO links is $c$, $2.998 \times 10^8$ m/s. We also assume that the data rate, $r$, for the RF link using Ka and L bands, and FSO, are different, denoted as $r_{ka}$, $r_{l}$, and $r_{o}$. To simplify the notations, the data rate on the Ka-band link is denoted as $r_{k}$. 

The distances between a GS as a SatNetOps Center and the source NGSO satellite and source GEO satellite are $d_0$ and $d_1$, respectively. The distance from a source GEO satellite to a target GEO satellite is $d_2$, and the distance from the destination GEO satellite to a destination LEO satellite is $d_4$. The distance from the source LEO satellite to the destination LEO satellite is $d_3$. Let the total path length of these hops be $L_{h}$. In the GLM scheme, $L_h = d_1 + d_2 + d_4$, and the LLM scheme, $L_h = d_0 + d_3$. Suppose the number of hops in a SatNetOps scheme is $n_{h}$. The latency measure can be expressed as follows:

\begin{equation}\label{Eq:latency}
 D =  \sum^{n_h}_{i=1}{\left( \frac{d(i)}{c} + M / r(i) + m + k \right)},
\end{equation}
where $d(i)$ and $r(i)$ are the distance and data rate of the $i$th hop.

The reliability measure is defined as the overall reliability of all links:
\begin{equation}
    \Phi = \left(1 - \prod_{i=1}^{n_h}{(1-\phi(i))} \right),
\end{equation}
where $\phi(i)$ is the reliability of a link on a path. 

In an NGSO satellite constellation with a homogeneous platform, we can assume the reliability of a link has the same reliability. The reliability can be modeled with multiple factors, such as satellite system dependability, mean-time-to-failure (MTTF), link stability, etc. Here we mainly consider the general factors affecting the link reliability from the communications perspective (e.g., propagation characteristics of RF and FSO signals, etc.) A GEO satellite usually has a longer design lifespan than an NGSO satellite. The GEO satellites' links (i.e., ground-to-space and space-to-space) are readily accessible. To these reasons, \hl{the reliability of a GEO satellite link is assumed to be higher than that of an NGSO link} \cite{Emily2018}, i.e., $\phi(i)_{GEO} > \phi(i)_{NGSO}$. The values used for $\phi(i)$ in Table \ref{Tbl:parameters} are also aligned with the commercial satellite constellation \cite{SES2022} where the service-level agreement network up-time and availability is $\ge 99.5\%$. Due to the different designs and deployment of an FSO system, the reliability between RF and FSO systems will not be compared. 

\begin{table}[ht]
\centering
\setlength{\tabcolsep}{0.5em} 
\renewcommand{\arraystretch}{0.95}
\caption{Evaluation Scenarios S1--S4}
\label{Tbl:eval_scenarios}
\begin{tabular}{c c c c c c}
\toprule
{} &
{\bfseries GS-LEO} & {\bfseries GS-GEO} & {\bfseries LEO-LEO} & {\bfseries GEO-GEO} & {\bfseries GEO-LEO}\\ 
\midrule
{\bfseries S1} & RF (Ka) & RF (Ka) & RF (Ka) & RF (Ka) & RF (Ka)\\
{\bfseries S2} & FSO & FSO & FSO & FSO & FSO\\
{\bfseries S3} & FSO & FSO & RF (Ka) & RF (Ka) & RF (Ka)\\
{\bfseries S4} & FSO & FSO & RF (Ka) & RF (Ka) & RF (L) \\

\bottomrule
\end{tabular}
\end{table} 

\begin{table}[ht]
\centering
\setlength{\tabcolsep}{0.5em} 
\renewcommand{\arraystretch}{0.95}
\caption{Key Simulation Parameters}
\label{Tbl:parameters}
\begin{tabular}{c r l}
\toprule
{\bfseries Parameter} & {\bfseries Value} & {\bfseries Notes}\\ 
\midrule
$r_{o,gl}$ & 1.8 Gbps \cite{Zech15} & Rate of the FSO GEO-LEO link \\
$r_{r,gl}$ & 324 Mbps \cite{Emily2018} & Rate of the RF GEO-LEO link \\
$r_{k}$ & 324 Mbps & Rate of the Ka-band link \\
$r_{l}$ & 150 kbps & Rate of of the L-band link \\
$M$ & \{512, 1024\} B & TF size of a TC packet\\
$T$ & 24 hr & Mission duration \\
$T_{start}$ & 2022-01-01 22:23:24 & Mission start date and time \\
$T_{end}$ & 2022-01-02 22:23:24 & Mission end date and time \\
$T_{sample}$ & 600 s & Sample time \\
$k$ & 100 $\mu$s & Avg. processing delay\\
$m$ & \{0, 100\} $\mu$s & Avg. queuing delay\\
$\phi_1(i)$ & 0.998 & Reliability of a LEO ISL\\
$\phi_2(i)$ & 0.999 & Reliability of a GEO ISL\\
$\phi_3(i)$ & 0.999 & Reliability of a GEO-LEO link\\

\bottomrule
\end{tabular}
\end{table} 

            

\section{Performance Evaluation}
\begin{figure}[!t]
\centering
\includegraphics[width=0.92\linewidth]{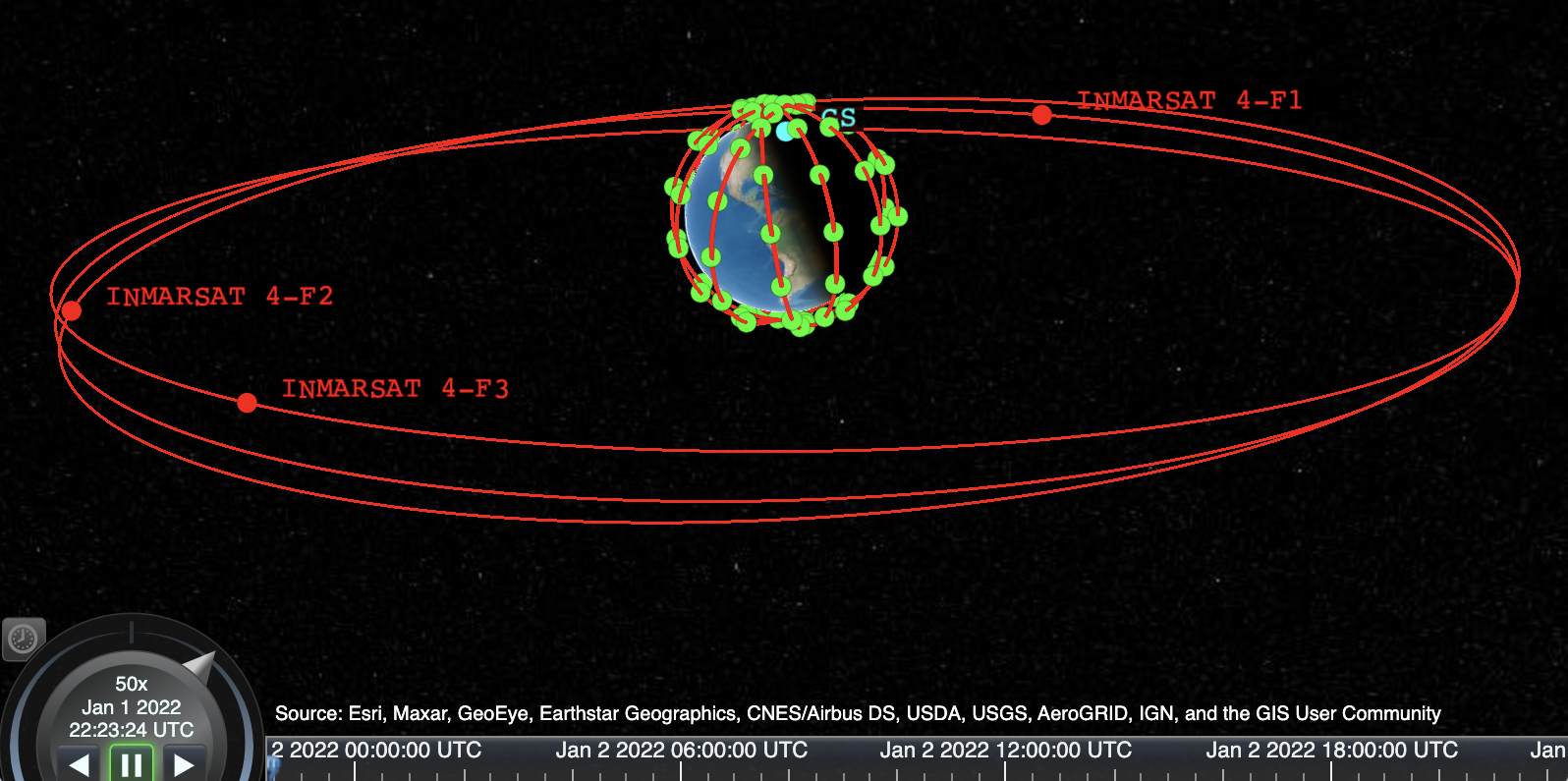}
\caption{Example satellite scenario}
\label{Fig:scenario_view}
\vspace{-15pt}
\end{figure}

The performance of the proposed SatNetOps TC missions is evaluated in MATLAB simulations based on a satellite scenario shown in Fig. \ref{Fig:scenario_view}. In the simulations, we generate the ephemeris data from the existing Inmarsat-4 GEO satellites and Telesat LEO satellites in polar orbit from the public filing. The GEO satellites ephemeris are propagated based on the public two-line element (TLE) data. The Telesat polar constellation can cover the polar regions with the optical payloads, where the inclination is 98.98$^{\circ}$ and altitude is 1015 km with 78 satellites in 6 orbital planes. The GS chosen is in Iqaluit, the capital city of Nunavut in northern Canada, with a minimum elevation angle of 30$^{\circ}$. To generalize the simulation scenarios, we consider the RF and FSO options on space-to-space and ground-to-space links in four typical scenarios, named S1--S4. These scenarios are displayed in Table \ref{Tbl:eval_scenarios}, where S4 considers the L-band satellite link for GLM considering the legacy satellite communication payloads; and RF (Ka) and RF (L) indicate a link uses Ka and L band, respectively.

The key simulation parameters are shown in Table \ref{Tbl:parameters}, where The data rates are based on the published results.  
Our simulations aim to obtain generalizable results. All LEO satellites in a constellation will get a chance to be a target satellite in iterations, and our results are averaged over all iterations over a 24-hr mission to obtain sufficient data. 

The assumptions we made in the simulations are based on the following justifications. We consider RF and optical payloads in this case. For the RF payload, we consider \hl{L-band} and Ka-band, and adopted the parameters reported in \cite{Zech15, Emily2018}. For the GLM solution, we assume the communication link(s) exist between GEO and LEO networks. The L-band has been widely available on legacy GEO/LEO satellites, and it can provide a good indication for the use of the S-band in some communications satellites. The Ka-band has been a popular option for recent satellites providing broadband access. Based on the CCSDS standards for space packet protocol (SPP) and TC-SDLP \cite{CCSDS133.0-B-2, CCSDS232.0-B-4}, an SPP packet can have a maximum length of 65542 B, and transfer frame (TF) in TC-SDLP has a maximum length of 1024 B. Thus, we let $M=\{512, 1024\}$ B in the experimentation. 

When transmitting the 64 KB SPP packet, there is a segmentation process where the packet will be split into multiple TFs with no re-transmission. This process is compatible with Type-A and Type-B services in TC-SDLP, as Type-A service allows re-transmissions but Type-B does not support it. For the processing delay per satellite note, we consider the parameter from the experimental study from \cite{Carlsson2004DelayPI}, where the mean delay for UDP/ICMP payload size ranging from 32 B to 1450 B is around 100 $\mu$s. Since these payload sizes match the values of $M$ for our simulations, we let $k=100 \mu$s. As for the queuing delay $m$, although it depends on various factors, such as traffic, buffer configuration, and algorithmic implementations, \hl{it is reasonable to assume zero to a slight latency on the same scale of $k$, in this case, $m=\{0, 100\}\mu$s.}


\begin{figure}[!t]
\centering
  \subfloat[\label{Fig:latencyS13}]{%
       \includegraphics[width=0.75\linewidth]{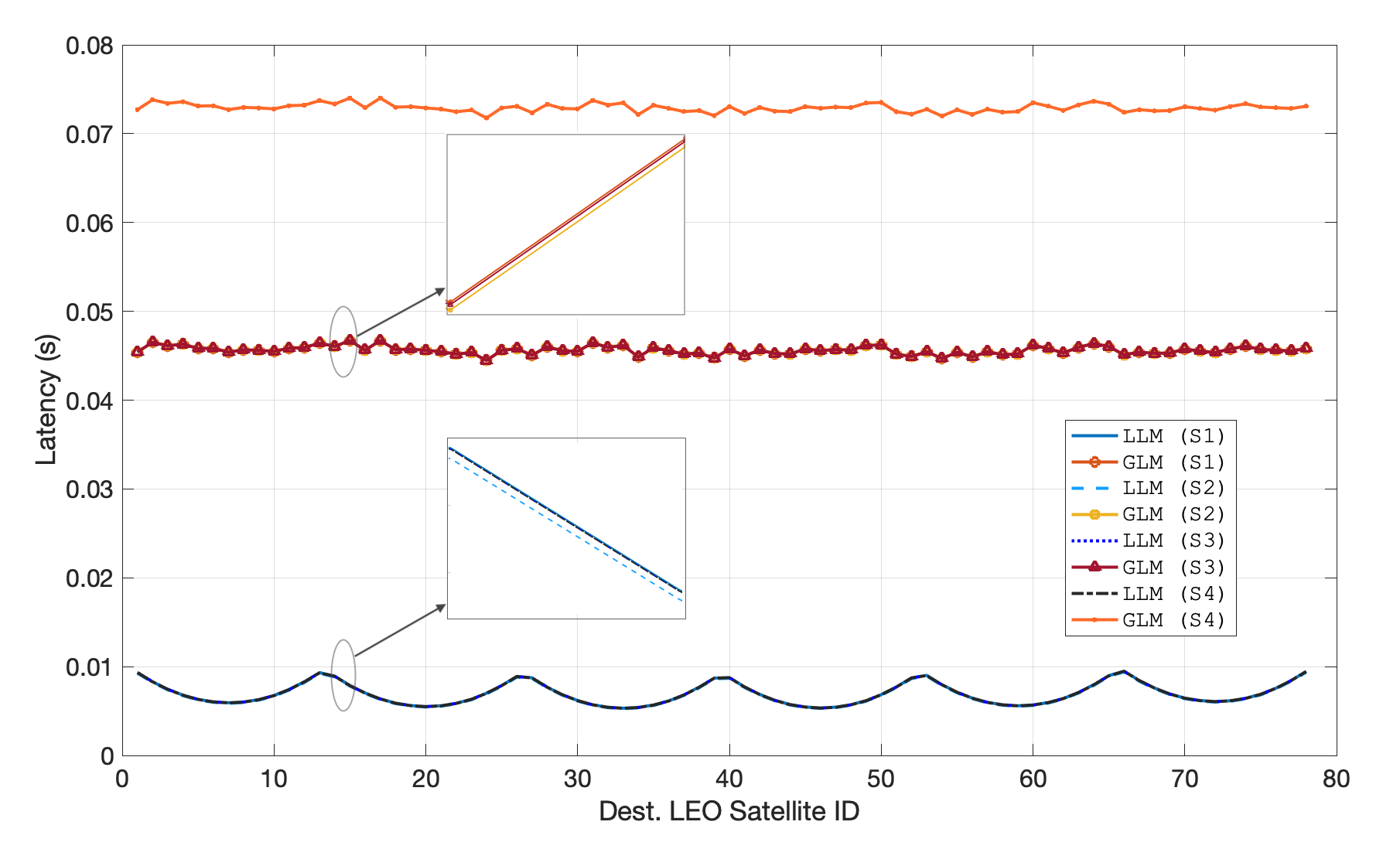}}
    \vfill 
  \subfloat[\label{Fig:latencyS4}]{%
        \includegraphics[width=0.75\linewidth]{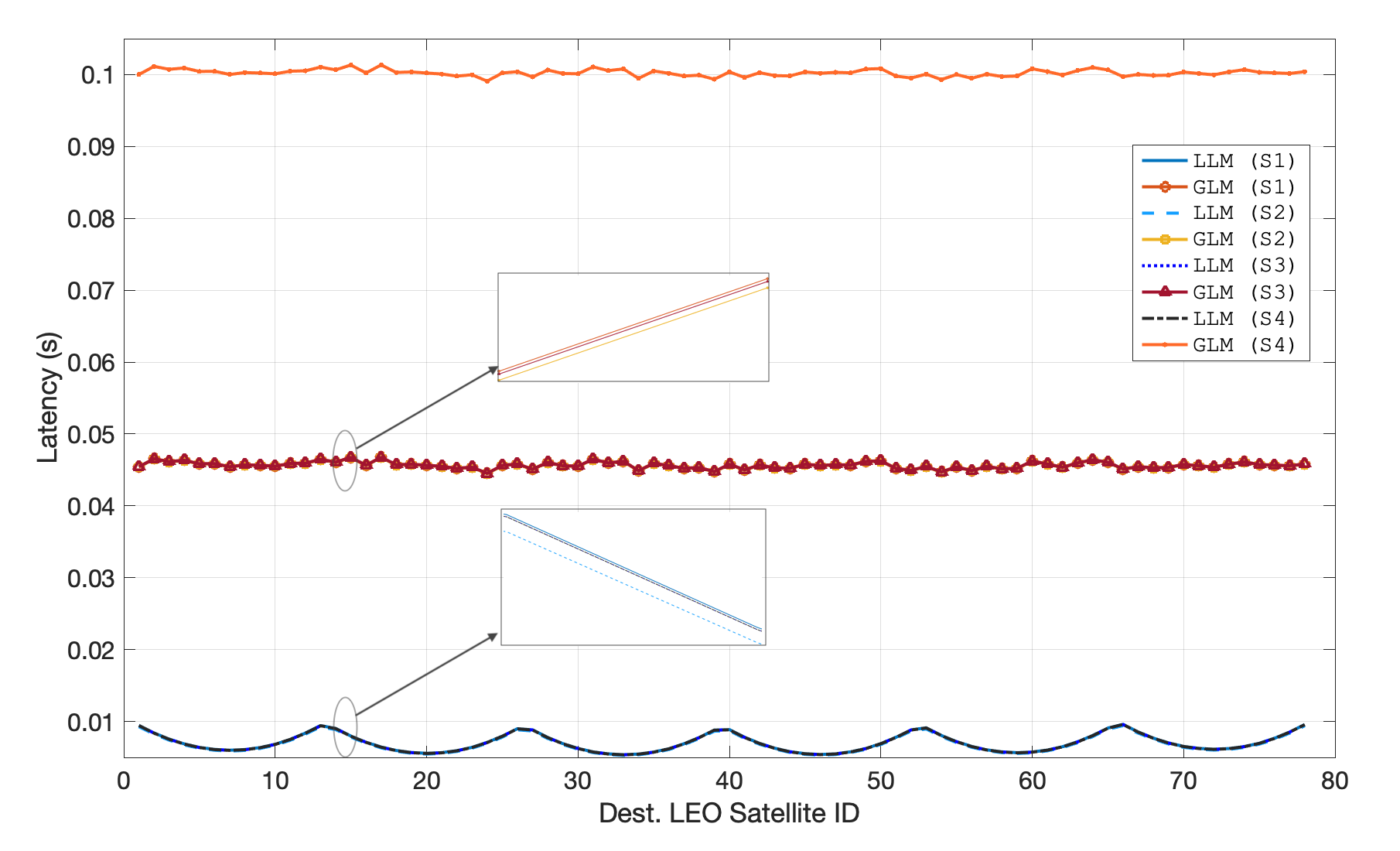}}
\caption{Latency performance in S1--S4 where $k=100\mu$s and $m=0\mu$s. (a) $M$=512 B (b) $M$=1024 B}
\label{Fig:latency}
\vspace{-15pt}
\end{figure}

\begin{figure}[!t]
\centering
\includegraphics[width=0.8\linewidth]{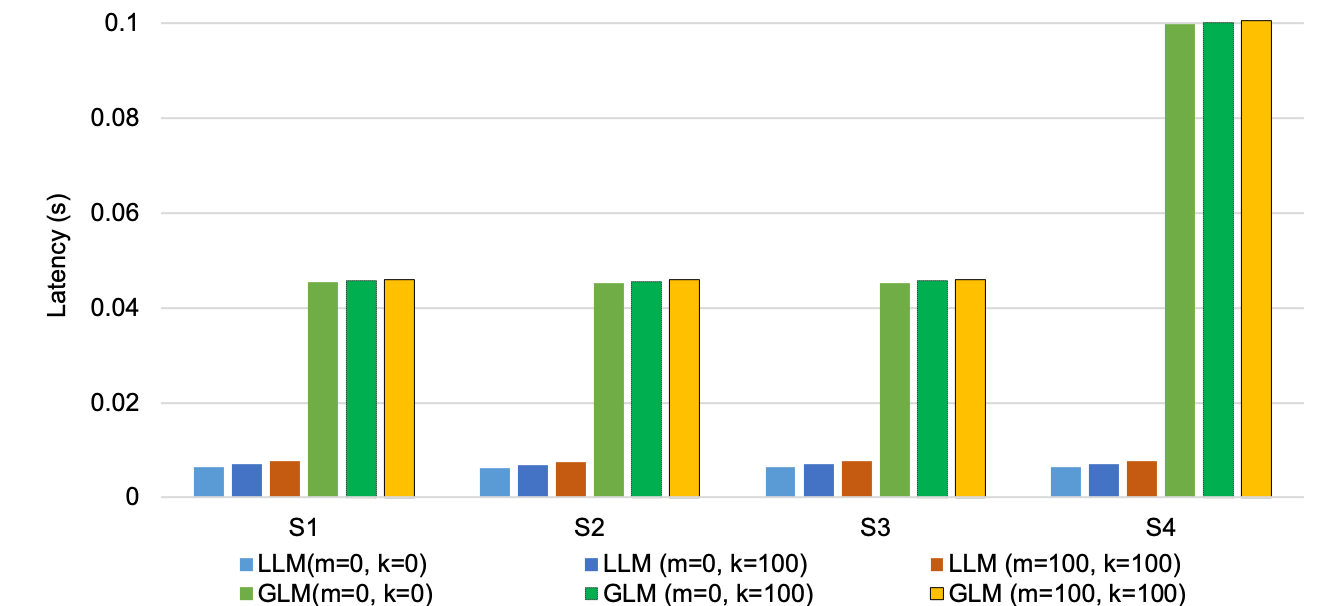}
\caption{Overall latency performance in S1--S4 ($M=1024$) for three cases: (1) $m=0, k=0$; (2) $m=0, k=100$; and (3) $m=100, k=100$}
\label{Fig:latency_allsats}
\end{figure}

\begin{figure}[!t]
\centering
  \subfloat[\label{Fig:latency3_a}]{%
       \includegraphics[width=0.76\linewidth]{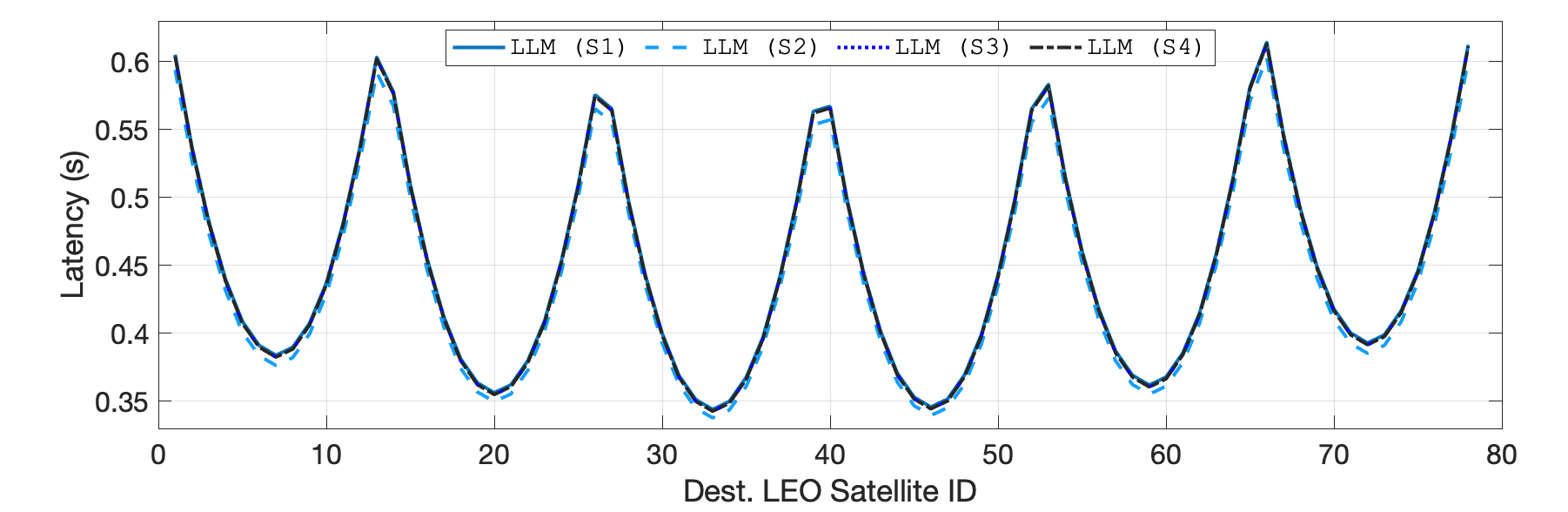}}
    \vfill 
  \subfloat[\label{Fig:latency3_b}]{%
        \includegraphics[width=0.76\linewidth]{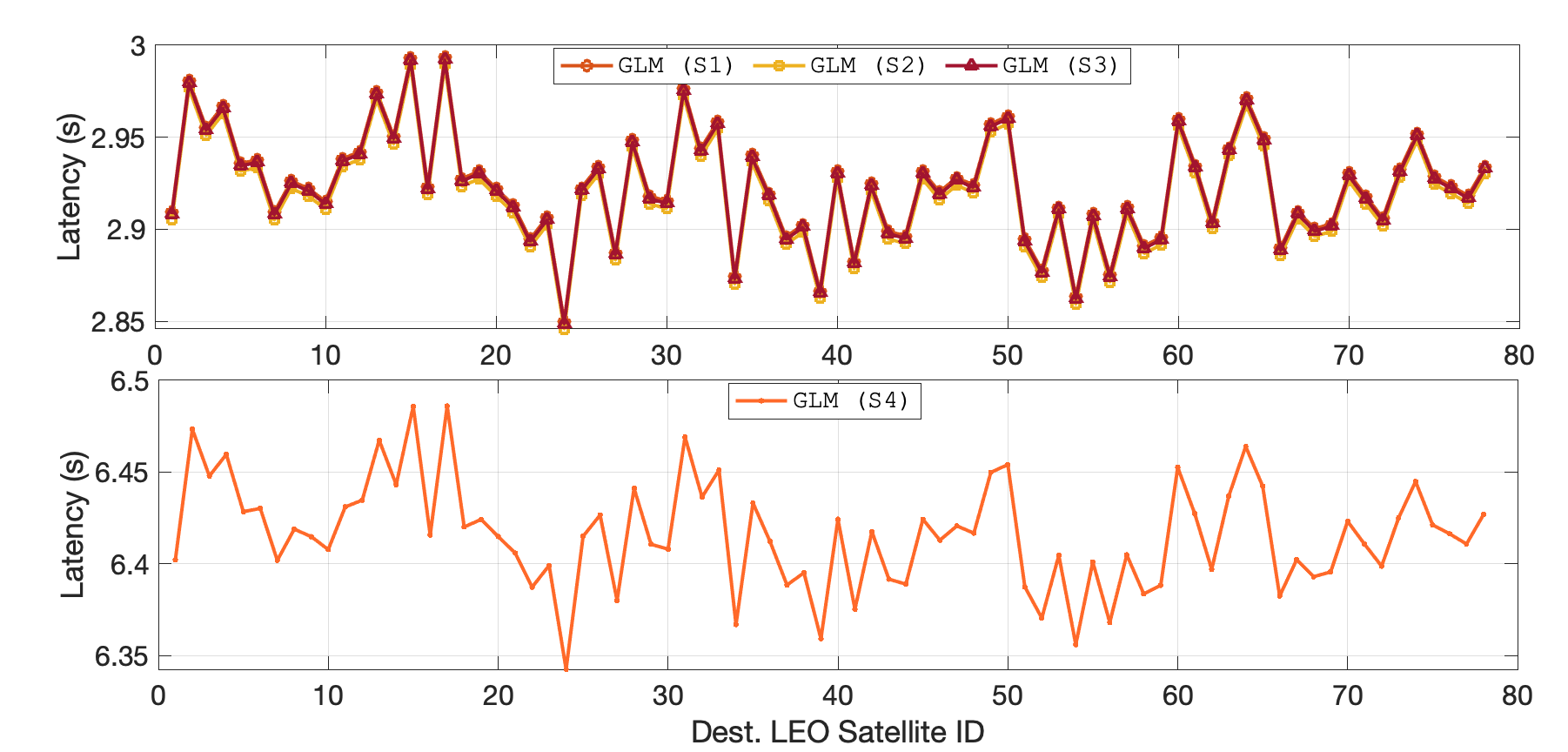}}
\caption{Latency performance in S1{--}S4 for an SPP packet transfer in (a) LLM and (b) GLM schemes where $k=100, m=0, M=1024$}
\label{Fig:latency_sp1}\vspace{-10pt}
\end{figure}

\subsection{Simulation Results}
\begin{figure}[!t]
\centering
\includegraphics[width=0.8\linewidth]{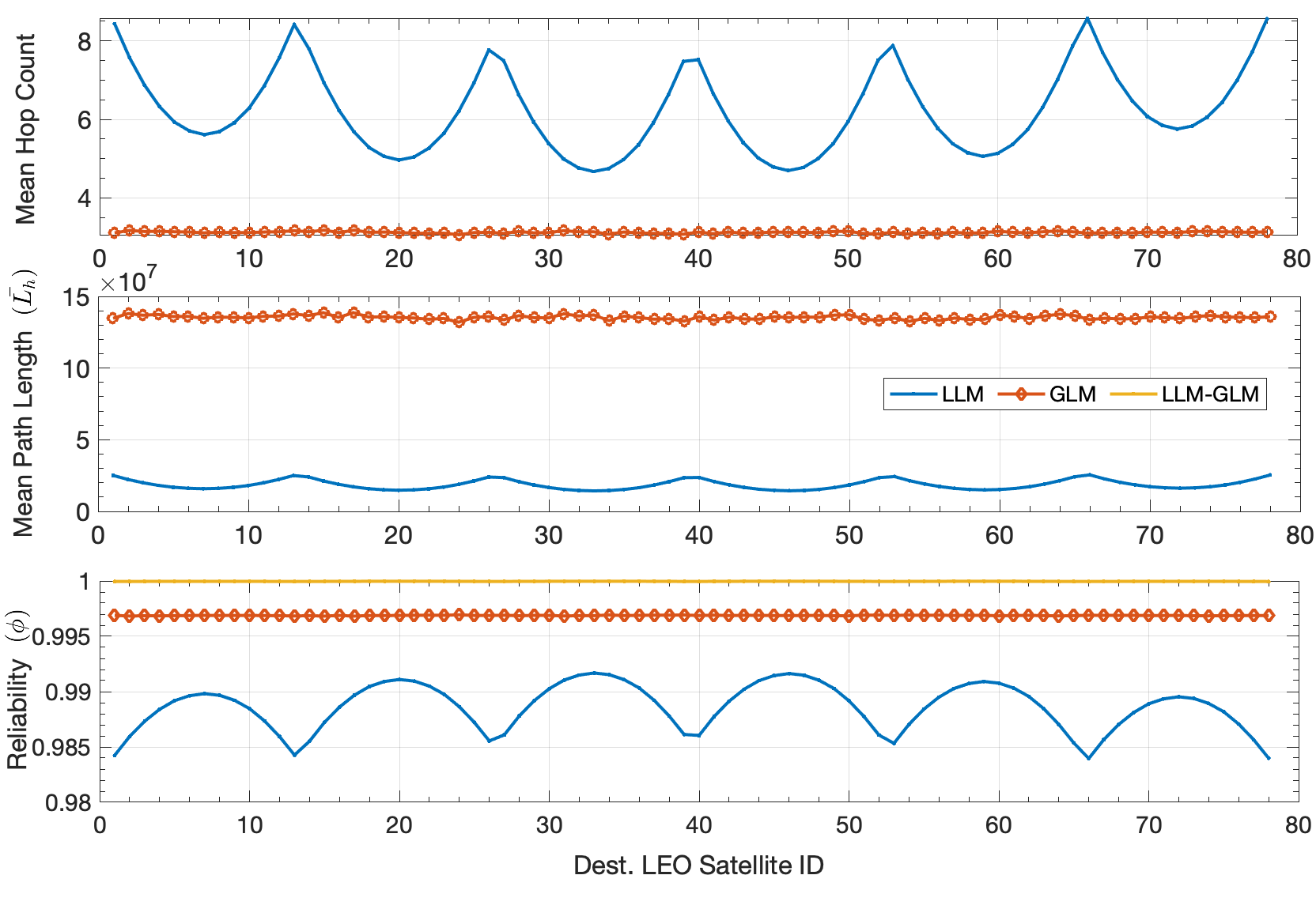}
\caption{Mean hop count, path length, and reliability in all scenarios}
\label{Fig:mean_path_len} \vspace{-15pt}
\end{figure}

As shown in Fig. \ref{Fig:latency}, when $M$=512 B, S2 has the lowest latency in LLM and GLM. The mean latency values for all destination LEO satellites of LLM and GLM are 6.9 ms and 45.6 ms, respectively. Due to the small size of the TC packet, the latency variations for LLM and GLM in S1--S3 are small. For GLM, the latency in S4 increases significantly due to the low data rate on the last GEO-LEO link, where the mean latency value is 72.9 ms. We can also see the latency in LLM has a correlation to the hop count (as shown in Fig. \ref{Fig:mean_path_len}). When $M$ is increased to 1024 B, LLM maintains its overall low latency compared to GLM, and S2 has the lowest latency in all scenarios, where its mean latency is similar to the case of $M=512$ due to the high data rate on the FSO links. The latency of GLM in S4 is increased to 100.2 ms due to the increased size of the TC packet. 

In Fig. \ref{Fig:latency_allsats}, the average latency performance for all satellites is shown, where we can see the case when $m=0, k=100$ has lower latency than the case when $m=100, k=100$. To demonstrate a lower bound we can achieve in the proposed schemes, we plot the case when $m=0, k=0$, indicating there is no process and queuing delays on satellite nodes, and this case shows the lowest latency than the previous two cases. 

\begin{figure}[!t]
\centering
\includegraphics[width=0.8\linewidth]{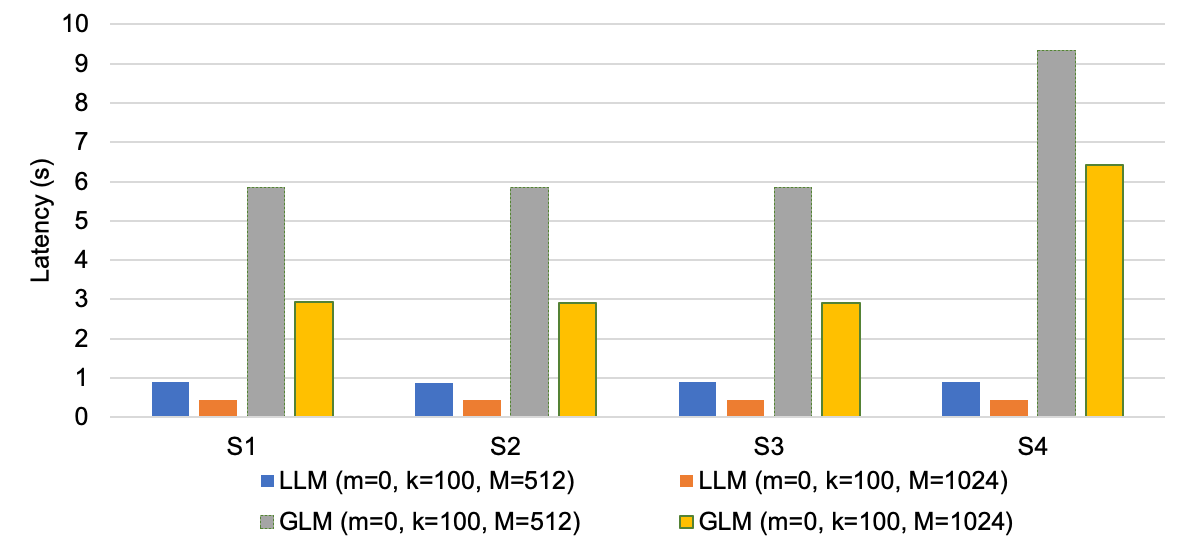}
\caption{Overall latency performance for an SPP packet transmission where $M=512$ and $M=1024$}
\label{Fig:latency_sp2}
\end{figure}

Now let us evaluate the scenario when an SPP packet with the size of 65542 B is transmitted. There is a segmentation process where the packet is split into TC packets subject to the value of $M$. In Fig. \ref{Fig:latency_sp1}, we can see LLM has the lowest latency than GLM. In Fig. \ref{Fig:latency_sp1}(b), two subplots show the latency values for S1--S3 and S4, respectively,  where the use of the L-band on the GEO-LEO link in S4 still results in the lowest latency performance. In Fig. \ref{Fig:latency_sp2}, the overall latency performance for all destination LEO satellites is shown for LLM and GLM, where we can see that the average latency decreases when $M$ increases due to less segmented packets. LLM in S2 has the lowest latency of 881.3 ms, and GLM in S4 has the worst latency of 9.3343 s. When $M=512$ B, GLM takes 6.624 to 10.51 times longer than LLM to transfer the SPP packet, while when $M=1024$ B, GLM takes 6.612 to 14.31 times longer than LLM to transmit the SPP packet.

Fig. \ref{Fig:mean_path_len} shows the mean hop count, mean path length, and reliability. We can see in Fig. \ref{Fig:mean_path_len}, the mean hop count and path length of GLM are steadier than those of LLM with regard to the destination LEO satellites. This is due to the constant GEO satellite availability to GS as their orbital period matches the Earth's rotation. The overall mean path length values of LLM and GLM are 1.8682e+07 m and 1.3539e+08 m, respectively. This means the path length of LLM is 7.2475 times shorter than GLM. The reliability performance is related to the type of links in the path based on the assumption that the reliability of an LEO-LEO ISL is slightly less than that of a GEO-LEO/GEO ISL. In addition, the maximum and minimum mean hop counts per destination LEO satellite in LLM are 8.5798 and 4.6723, respectively. For all events, the maximum and minimum hop counts are 17 and 0, respectively, where 0 hop indicates the GS can directly contact the destination satellite. The hop count is expected to increase for a large constellation of LEO satellites due to the shorter distance between satellites. To see the extended scheme where LLM and GLM are used in parallel, referred to as ``LLM-GLM'' in the reliability plot, we can see the reliability can be significantly increased to over 99.99\%. 

\begin{figure}[!t]
\centering
\includegraphics[width=0.76\linewidth]{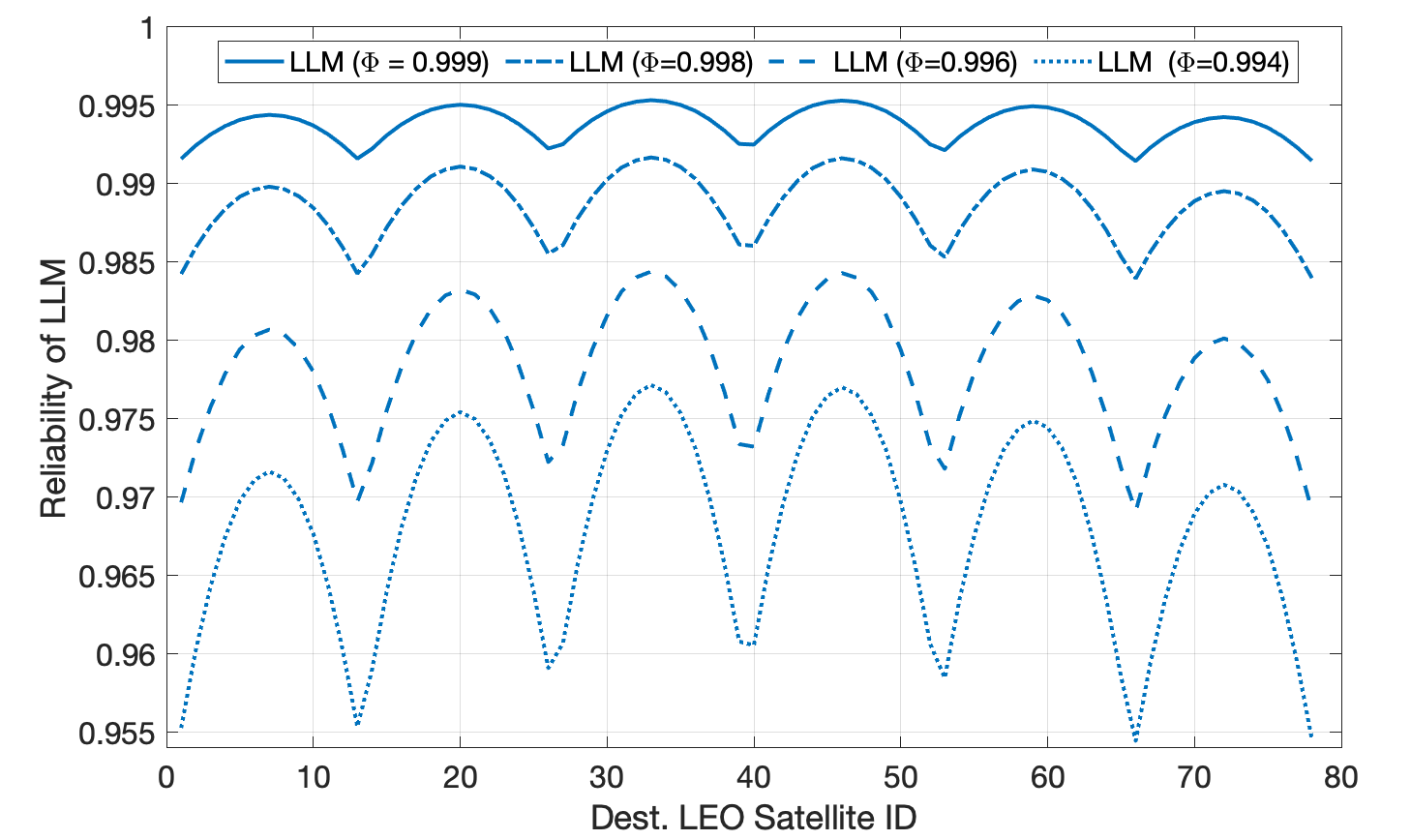}
\caption{Reliability of LLM when $\phi_1(i)$ decreases}
\vspace{-10pt}
\label{Fig:reliability_LLM}
\end{figure}

Fig. \ref{Fig:reliability_LLM} shows how $\phi_1(i)$ changes the overall reliability of LLM. The results in Fig. \ref{Fig:reliability_LLM} indicate that although LLM can reduce the latency compared to GLM, the cost is the overall reliability, which may result in the transmission impairments that may lead to the need for additional mechanisms for mitigation.

\subsection{Discussion of the Results}
To make the proposed SatNetOps schemes work, we need to ensure access status between a GS and the satellites. The probability of contact between a GS and a direct satellite determines the chances of a successful scheme execution. When evaluating the LLM scheme, there are 11310 data entries where we noticed that there are 2028 occurrences that GS has no immediate contact with LEO satellites. These occurrences are ruled out in data analysis to have a fair comparison between the GLM and LLM schemes. 
Although we used the GS in the northern region where other LEO constellations may have poor or no coverage, our proposed schemes can work in different satellite constellations (such as another LEO/MEO constellation) and GS locations. 

For compatibility with the heterogeneous payload and configurations on satellites, a relatively conservative data rate for Ka-band links is adopted in our evaluation. We noticed there the data rate can be higher, e.g., 600 Mbps, as reported in \cite{Sedin20}, where our results can still be used as an performance indication of the proposed schemes in this case. Subject to a specific configuration for an application, e.g., beams configuration and a coding \& modulating scheme, a fine-grained performance can be further derived. Such fine-grained results may be generated on a specific FSO setup due to the currently non-standard configurations on different satellite platforms.

The selection of $\phi$ in a real network will also have an impact on the reliability. When modeling the reliability with the considerations of system-level component dependability, a lower bound performance result may be obtained. The results shown in Fig. \ref{Fig:reliability_LLM} also indicate the additional measures that may be required to compensate for the cost of reduced reliability for LLM. This also indicates that the hop count on a path for a TC message transfer should be maintained at a reasonable level. In a larger LEO satellite constellation than the one used in experimentation, when a path contains many hops, the reliability may be reduced and the $k$ and $m$ on the satellite nodes may further introduce latency in the LLM scheme. Therefore, there is a trade-off in an LLM scheme that requires careful analysis before tailoring the LLM parameters for a specific mission and LEO satellite constellation.

For the generality, the single flow scenario is considered in our evaluation. Our results can be extended to multi-flow scenarios subject to the implementations considering RF and FSO for space-to-space and ground-to-space links. In addition, extra delays may be introduced if access and specific routing schemes are employed. 



\section{Conclusion}
The proposed SatNetOps approach provides a new way of addressing the increasing operations challenges imposed by the upcoming NGSO satellite constellations. This paper validates the effectiveness of the proposed approach with two feasible schemes. These schemes can be applied or extended to other scenarios for TM/TC and network management missions where timing and reliability performance needs to be assured. There is still much room for future contributions. For example, using RF/FSO channel models, data rates subject to coding and modulation scheme, and additional scenarios using different NGSO satellite constellations will be explored in future work.


%

\section*{Acknowledgment}
This work was partially supported by the High-Throughput and Secure Networks Challenge program of National Research Council Canada. We also acknowledge the support of the Natural Sciences and Engineering Research Council of Canada (NSERC), [funding reference number RGPIN-2022-03364].


\ifCLASSOPTIONcaptionsoff
  \newpage
\fi



\bibliographystyle{IEEEtran}
\bibliography{./references}
%



%



\begin{IEEEbiographynophoto}
{Peng Hu}
received his Ph.D. degree in Electrical Engineering from Queen's University, Canada. He is currently a Research Officer at the National Research Council Canada and an Adjunct Professor at the Cheriton School of Computer Science at the University of Waterloo. He has served as the Technology Working Group Co-Chair of the IEEE LEO Satellites \& Systems (SatS) project, an Associate Editor of the IEEE Canadian Journal of Electrical and Computer Engineering, a voting member of the IEEE Sensors Standards committee, and on the organizing/technical committees of industry consortia and international conferences/workshops at IEEE ICC'23, IEEE GLOBECOM'21, IEEE PIMRC'17, IEEE AINA'15, etc. His current research interests include satellite-terrestrial integrated networks, autonomous networking, and industrial Internet of Things systems.

\end{IEEEbiographynophoto}



\end{document}